\documentclass[aps,prb,reprint,longbibliography,superscriptaddress]{revtex4-1}

 \usepackage{amsmath,bm}
 \usepackage{mathrsfs}
 \usepackage{amsfonts}
 \usepackage{soul}
 \usepackage{setspace} 
 \usepackage{graphicx}
 \usepackage{epstopdf}
 \usepackage{dcolumn}
 \usepackage{amsmath}
 \usepackage{epsfig}
 \usepackage{indentfirst}
 \usepackage{psfrag}
 \usepackage{subfigure}
 \usepackage{amssymb}
 \usepackage{color}
 \usepackage{units} 
\usepackage{physics}
\usepackage{dcolumn}
\usepackage{bm}
\usepackage{bbold}
 \usepackage{dcolumn}
 \usepackage{natbib}
 \usepackage{wrapfig}
 \usepackage{xcolor}

\usepackage[backref=none,bookmarksnumbered=true,bookmarks=true,bookmarksopen=true,colorlinks=true,
citecolor=blue,linkcolor=blue,anchorcolor=green,urlcolor=blue,unicode=false]{hyperref}

\newcommand{\bipd}{$\beta$-Bi\textsubscript{2}Pd }
\newcommand{\bipdp}{$\beta$-Bi\textsubscript{2}Pd}
\newcommand{\sqrtfive}{$\mathrm{\sqrt{5}a}$ }
\newcommand{\sqrttwo}{$\mathrm{2\sqrt{2}a}$ }
\newcommand{\adir}{2a-(100) }
\newcommand{\sqrtfivedir}{$\mathrm{\sqrt{5}a-(120)}$ }

\newcommand{\zeroysr}{$A^{0}$ }
\newcommand{\oneysr}{$A^{1}$ }

\usepackage{ulem}[normalem] 

\makeatletter
\newcommand\colorsout[1]{\bgroup \markoverwith{\textcolor{#1}{\rule[0.5ex]{2pt}{0.4pt}}}\ULon}

\makeatother

\AtBeginDocument{%
    \newwrite\bibnotes
    \def\bibnotesext{Notes.bib}
    \immediate\openout\bibnotes=\jobname\bibnotesext
    \immediate\write\bibnotes{@CONTROL{REVTEX41Control}}
    \immediate\write\bibnotes{@CONTROL{%
    apsrev41Control,author="08",editor="1",pages="1",title="0",year="1"}}
     \if@filesw
     \immediate\write\@auxout{\string\citation{apsrev41Control}}%
    \fi
}%
\begin{document}

\title{Diluted Yu-Shiba-Rusinov arrays on the \bipd anisotropic superconductor}

\author{Stefano Trivini}
  \affiliation{Centro de Física de Materiales (CFM-MPC), Centro Mixto CSIC-UPV/EHU, 20018 San Sebastián, Spain}
  \affiliation{CIC nanoGUNE-BRTA, 20018 Donostia-San Sebasti\'an, Spain}

  \author{Jon Ortuzar}
  \affiliation{CIC nanoGUNE-BRTA, 20018 Donostia-San Sebasti\'an, Spain}

  \author{Javier Zaldivar}
  \affiliation{CIC nanoGUNE-BRTA, 20018 Donostia-San Sebasti\'an, Spain}

\author{Edwin Herrera}
  \affiliation{Departamento de Física de la Materia Condensada, Instituto Nicolás Cabrera and Condensed Matter Physics Center (IFIMAC), Universidad Autónoma de Madrid, Madrid, Spain }

\author{Isabel Guillamón}
  \affiliation{Departamento de Física de la Materia Condensada, Instituto Nicolás Cabrera and Condensed Matter Physics Center (IFIMAC), Universidad Autónoma de Madrid, Madrid, Spain }
  
\author{Hermann Suderow }
  \affiliation{Departamento de Física de la Materia Condensada, Instituto Nicolás Cabrera and Condensed Matter Physics Center (IFIMAC), Universidad Autónoma de Madrid, Madrid, Spain }

\author{F. Sebastian Bergeret}
    \affiliation{Centro de Física de Materiales (CFM-MPC), Centro Mixto CSIC-UPV/EHU, 20018 San Sebastián, Spain}
	\affiliation{Donostia International Physics Center (DIPC), 20018 Donostia-San Sebasti\'an, Spain}
  
\author{Jose Ignacio Pascual}
  \affiliation{CIC nanoGUNE-BRTA, 20018 Donostia-San Sebasti\'an, Spain}
\affiliation{Ikerbasque, Basque Foundation for Science, 48013 Bilbao, Spain}

\begin{abstract}
Magnetic adatoms on s-wave superconductors induce bound states inside the superconducting gap, called Yu-Shiba-Rusinov states (YSR). The anisotropy of the Fermi surface determines the spatial extension of bound states in a quasi-two-dimensional superconductor. This is especially important in the diluted impurity limit since the orbital overlap determines the coupling of YSR states of neighboring atoms and the formation of the collective YSR system.  Here, we build diluted arrays of Mn atoms with different dimensionalities on the surface of \bipdp, and we measure the evolution of their YSR spectra with the structure. We detect the coupling as a split of YSR peaks in subgap spectra and find that the split size increases with the number of atoms. The orientation of the structures along different directions of the \bipd substrate modulates the split and particle-hole asymmetry of the YSR states due to the anisotropic character of the Fermi surface, captured by the Green function model. With the aid of the model, we found multiple YSR excitations in an extended 2D array of 25 Mn atoms, and we identified that their spatial distribution reflects a chiral LDOS. 

\end{abstract}
\maketitle

\section{Introduction}

The sub-gap Yu-Shiba-Rusinov (YSR) states are bound states for quasiparticles in superconductors localized on atomic or molecular spins. These are rich fingerprints of the exchange interaction between a magnetic impurity and a superconductor. YSR excitations can be detected with scanning tunneling spectroscopy (STS) as narrow resonances inside the superconducting gap \cite{franke2011,heinrich2018}. YSR resonances can be spin-polarized \cite{cornils2017,wang2021} and are described with classical \cite{yu1965,shiba1968,rusinov1969} or quantum spin models \cite{zitko2008,vonoppen2021}. Their multiplicity reflects the multiple orbitals of the atoms coupled to the conduction electrons \cite{ji2008,ruby2016}, the magnetic anisotropy YSR split components \cite{hatter2015,trivini2023} or the coupling between impurities that results in symmetric and anti-symmetric states \cite{choi2018,friedrich2021,kezilebieke2018,ruby2018,ding2021,schmid2022}.

\begin{figure}[b!]
\begin{centering}
			\includegraphics[width=1\columnwidth]{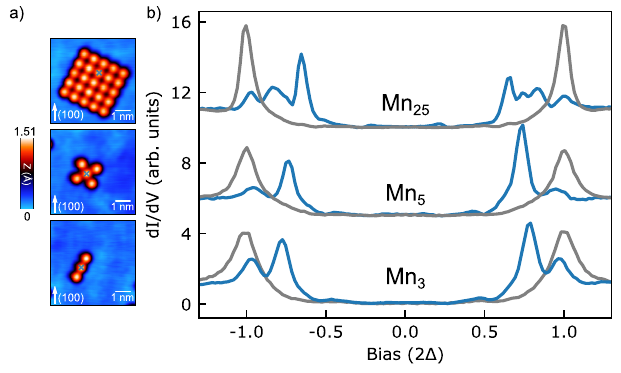}    
			\caption{\textbf{Comparison of YSR excitation for increasing dimensionality.} \textbf{a)} The topography of three Mn structures with period \sqrtfive $\sim 0.75$ nm (being $a$ the ab-plane constant lattice of the $\beta$-Bi$_2$Pd) and with 3,5 and 25 Mn atoms (V = 3 mV, I = 100 pA). On the right, Differential conductance spectra measured with a superconducting tip on the structures in the atom indicated by the blue cross (blue line) and on the superconductor as a reference (gray line), showing a gap closing and multiple YSR components rising (V = 3 mV, I = 300 pA, bias normalized to $2 \Delta$ units for easier comparison).}
		\label{F1}
  \end{centering}
		\end{figure}

The formation of YSR bands has been found in 1D \cite{schneider2022,schneider2021} and 2D \cite{kezilebieke2020,manna2020,soldini2023a} densely packed atomic and molecular lattices \cite{farinacci2023,vano2023}. In ensembles of diluted impurities, coupling between YSR states is indirect, therefore the pattern of YSR states is determined by the  Fermi surface of the superconducting host \cite{flatte2000,morr2003,menard2015,liebhaber2022,kim2020,uldemolins2022,ortuzar2022}. Studies of diluted 2D atomic lattices are still lacking, despite their potential for topological superconducting phases \cite{rontynen2015,li2016}. 

Here, we build 1D and 2D YSR arrays of paramagnetic Mn impurities on the \bipd superconductor surface by lateral manipulation using the tip of an STM and study the evolution of the YSR resonances in various configurations. Due to the anisotropy of the \bipd Fermi surface, the strength of the YSR coupling varies along the different crystallographic directions. Using STS, we detect both energy shifts of the YSR states and multiple components arising [Fig.~\ref{F1}b] due to the YSR overlap. 

We use the classical spin model of \cite{ortuzar2022} to develop our own code for simulating the local DOS of arbitrary N-atom arrays \cite{green-ysr}. The local DOS is governed by the exchange coupling, the Fermi surface dimensionality, the inter-atomic spacing, and the spin texture, all parameters that can be implemented in the model. The advantage of the exact Green Functions is that we can calculate the local DOS and its anisotropic character, accounting analytically for the non-circular shape of the Fermi contour of the \bipdp. 

The article is organized in the following way. In section II, we introduce the spectral properties of single Mn atoms on \bipdp, and the observation of split peaks in dimers as a fingerprint of their interaction. In section III, we study the evolution of the Mn YSR states in ordered clusters of 5 atoms as a function of their orientation on the \bipd lattice. In section IV, we report the evolution of YSR bands in linear chains. In section V, we finally present a two-dimensional structure formed by 25 Mn atoms and identify the appearance of collective YSR modes. 

\section{Mn SINGLE ATOMS AND DIMERS ON \bipd}

\begin{figure}[t]
\begin{center}
			\includegraphics[width=1\columnwidth]{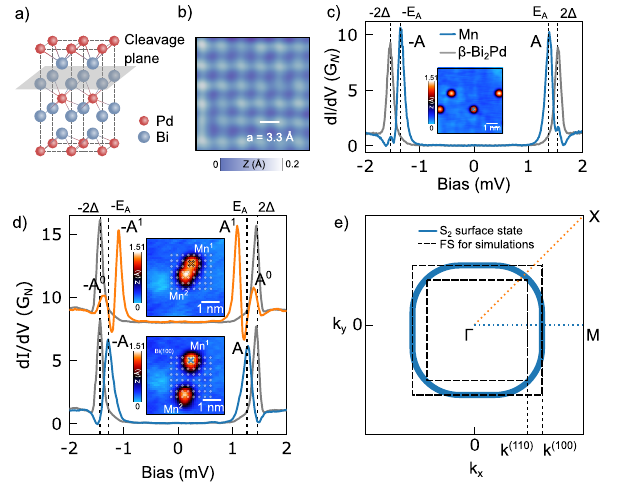}    
			\caption{\textbf{Mn on \bipd superconductor.} \textbf{a)} Crystal structure of \bipd. \textbf{b)} Atomic resolution of the Bi terminated surface (V = 3 mV, I = 10 nA). \textbf{c)} dI/dV spectrum of the bare \bipd surface (gray curve) and of a Mn atom (blue curve), using a $\beta$-Bi$_2$Pd tip. (inset) topography of the Mn atom (V = 3 mV, I = 35 pA). \textbf{d)} YSR spectrum of a Mn before (bottom) and after (top) forming an interacting Mn dimer, as shown in topography images  in the insets  (V = 3 mV, I = 300 pA). The reference spectrum of the \bipdp is shown in grey line, for comparison. \textbf{e)} Sketch of the S2 surface state's Fermi contour of the \bipd surface. In the simulations, we approximate it to a perfect square (outer dashed square) and use a lower $k_F$ along diagonal directions (e.g. the 110 as shown in the sketch) to account for the finite curvature along the (110) direction \cite{sakano2015,iwaya2017}.   
   }
		\label{F2}
  \end{center}
\end{figure}

The s-wave superconductor \bipd has a critical temperature $T_c=5.4$ K \cite{imai2012} and a superconducting gap $\Delta$ $\sim$ 0.75 meV \cite{herrera2015,che2016,kacmarcik2016}. The samples are prepared via mechanical exfoliation thanks to the directional Bi-Pd covalent bonding and the weak coupling between consecutive Bi-Pd-Bi trilayers of the \bipd structure [Fig.~\ref{F2}a] \cite{wang2017a}. This process exposes the square-symmetrical Bi-terminated surface with a lattice constant $a = 3.3$~\AA\ [Fig.~\ref{F2}b]. The Mn atoms are evaporated on a freshly cleaved sample placed at the cold stage of our STM (1.3K). The STM and STS measurements are done at a base temperature of 1.3 K and using a superconducting \bipd tip to enhance the energy resolution.

Manganese single atoms on the \bipd surface appear as 140 pm high protrusions in STM images. They are adsorbed on hollow sites among 4 Bi atoms,  as previously observed for Cr impurities \cite{choi2018}.  Figure~\ref{F2}c compares dI/dV spectra measured on an isolated Mn atom and on the \bipd surface nearby as a reference. The surface shows sharp quasiparticle peaks at $\pm$1.5 mV (i.e. at $\pm 2 \Delta$/e), as expected for the tunneling spectrum between superconducting \bipd tip and surface. The spectrum on the Mn atom also shows weak quasiparticle peaks at $\pm$1.5 mV and a sharp single YSR resonance at $\epsilon \approx\pm1.35$ mV, with similar weight in particle and hole components, which we label $A$.  Despite Mn being an S = 5/2 impurity, the observation of a single YSR resonance points towards a single YSR-channel configuration. Although we cannot discard the fact that additional degenerate orbital channels exist within that single YSR spectral peak, modeling YSR spectra with a single channel can majorly reproduce simulated spectra, as shown in the following sections. 

\begin{figure*}[t]
\begin{center}
    			\includegraphics[width=1\textwidth]{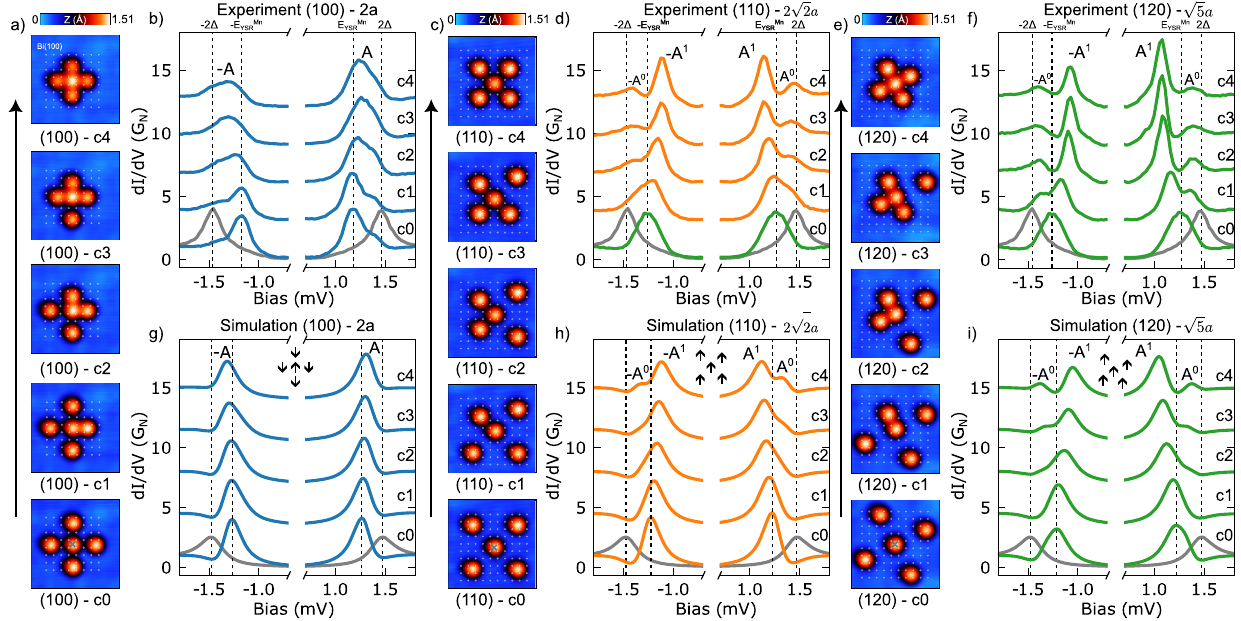}    
			\caption{\textbf{Mn$_5$ structures along different directions.} \textbf{a, c, e)} From bottom to top are the topography of the manipulation sequence to construct Mn$_5$ crosses along (100) from 3a to 2a, (110) from $4\sqrt{2}a$ to $2\sqrt{2}a$, and (120) from $2\sqrt{5}a$ to $\sqrt{5}a$ with $a = 3.3$~\AA\. (V = -100 mV, I = 40 pA). \textbf{b, e, h)} dI/dV spectra of the central atom of the Mn structures after moving each lateral Mn atom, offset vertically for clarity (V = 5 mV, I = 500 pA).  In gray the reference spectrum in \bipdp. The spectrum for stage (110) - c0 is missing, we use the (120) - c0 as a reference (at $2\sqrt{5}a$ and $3\sqrt{2}a$ the Mn atom are not interacting). \textbf{c, f, i)} Model simulation of the 5 impurity structures with the Green function approach. The spin orientation angle that is 90° for \adir while 0° for the other two configurations (simulation parameters $p_{F}^{100} = 0.32/a_0$, $p_{F}^{110} = 0.21/a_0$, $p_{F}^{120} = 0.28/a_0$, $m^* = 5.85/m_e$, $\alpha = \pi \nu_0JS/2 = 0.030$, $\beta = \pi \nu_0 U = 0.011$ where $a_0$ is the Bohr radius and $\nu_0$ is the normal metal density of states).
   }
		\label{F3}
  \end{center}
		\end{figure*}

As shown elsewhere \cite{liebhaber2022,soldini2023a}, in the dilute limit, magnetic adatoms can also magnetically couple via indirect exchange interactions, namely represented as finite overlap of their YSR wavefunctions \cite{flatte2000,morr2003}. The interaction can be detected in tunneling spectra as a peak split and/or shift varying with the adatom separation and relative alignment of the spins.
For example, in Fig.~\ref{F2}d, we show that the dI/dV spectrum on the Mn$^1$ adatom of the inset splits into two components \zeroysr and \oneysr when the Mn$^2$ adatom is laterally displaced with the STM tip to lay at the closest distance along the (120) surface direction, from  $\sim$ 15~\AA~to $\sim$ 7.5 \AA, forming a \sqrtfive dimer. Since the separation is larger than one unit cell of the \bipd surface, the adatoms do not interact directly nor via superexchange-like interactions \cite{choi2018}. Therefore, we attribute this YSR split to the overlapping of YSR states \cite{flatte2000}. 

Albeit split YSR peaks also reflect the presence of spin-orbit coupling \cite{beck2021} and, thus, delivers information about the pairing symmetry of the superconductor \cite{kim2015}, a crucial aspect to account for first is the anisotropy of the surfaces's bands. 
The \bipd superconductor has two surface states (S1 and S2) and three bulk bands crossing the Fermi level \cite{sakano2015,iwaya2017}, and all of them have a square shape. 
We thus consider in the following simulations a surface band with a square shape as sketched in Fig.~\ref{F2}e.
As shown in our previous work \cite{ortuzar2022}, the flat portions of the Fermi contour focus the YSR wavefunction along the low symmetry directions. Therefore,  the alignment between the atoms with respect to the high-symmetry substrate's directions is expected to crucially affect the type and strength of exchange interactions between YSR states.

\section{YSR Coupling Between Atoms Along Different Crystalline Directions}

The marked anisotropy of the substrate's density of states affects the pattern of YSR interactions in atomic ensembles. To demonstrate this effect, we build up a model atomic ensemble using lateral tip manipulation, consisting of a central adatom surrounded by four neighbors, i.e., an Mn$_5$ atomic array in a cross-like shape. This can be considered a building block for the 2D growth of atomic systems. Using dI/dV spectroscopy, we compared in Fig.~\ref{F3} the effect of the different arrangements of the Mn$_5$ structure on the central Mn adatom.  

When the adatoms are positioned at three unit cells away from the central Mn adatom along the (100) direction, the dI/dV spectrum on the central one is similar to the reference Mn spectrum in Fig.~\ref{F1}c, namely a single YSR state at $\epsilon \approx \pm 1.3$ mV [Fig.~\ref{F2}b]. 
Fig.~\ref{F2}a shows the sequence of atomic manipulation steps forming a compact Mn$_5$ structure oriented along the (100) direction with inter-atomic distance 2$a$ (i.e. 6.6 \AA). After every manipulation step, we probed the effect on the central adatom with dI/dV spectra to follow the gradual effect of activating YSR coupling along this direction. The results, shown in  Fig.~\ref{F3}b, reveal that the YSR resonance does not split and, instead, shows a small energy shift towards the gap edge, i.e. reduces the YSR binding energy. This is a sign of AFM coupling, which reduces the total spin of the Mn interacting system \cite{choi2017}. 
This behavior differs when the Mn$_5$ cluster is constructed instead along the (110) direction  [Fig.~\ref{F3}c]. As reported in Fig.~\ref{F3}d,  approaching the four adatoms to the closest distance \sqrttwo (i.e. 9.3 \AA) induces the splitting of the YSR resonance to form two components, \zeroysr and \oneysr. The lower peak \oneysr shifts gradually to lower energy with the number of adatoms added, and thus, the split between the two components increases. Repeating the atomic manipulation steps now along the (120) direction [Fig.~\ref{F3}e] to the form Mn$_5$ structure with closest distance $\sqrt{5}a$ (i.e. 7,4 \AA) also causes YSR splitting, now with even larger energy difference between the two components [Fig.~\ref{F3}f]. The observation of a large YSR splitting in the last two structures proves that the Mn adatoms in these orientations tend to align with parallel-like orientations, probably in a non-collinear fashion due to the strong spin-orbit cooling of this substrate \cite{mier2021a}.

\begin{figure*}[t]
\begin{center}
    			\includegraphics[width=1\textwidth]{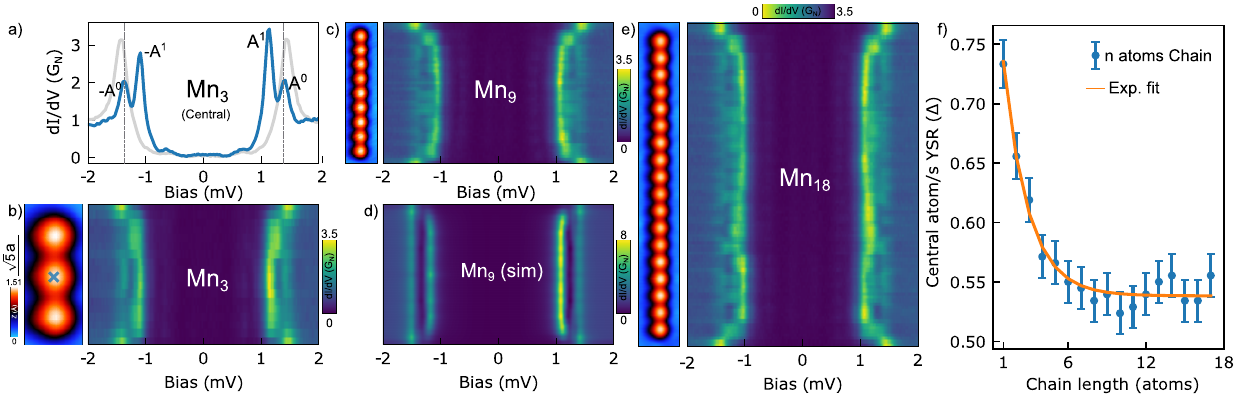}    
			\caption{\textbf{Mn$_{18}$ chain along (120).} \textbf{a)} Differential conductance spectra measured in the central atom of the Mn$_3$ chain and the reference on \bipd (V = 3 mV, I = 300 pA). \textbf{b)} Topography, on the left, and line of dI/dV spectra, on the right, measured along the Mn$_3$ chain (V = 3 mV, I = 300 pA). \textbf{c)} Topography and dI/dV for the Mn$_9$ chain. \textbf{d)} Simulation with our Green function model of the Mn$_9$ chain (simulation parameters: $p_{F}^{120} = 0.28/a_0$, $m^* = 5.85/m_e$, $\alpha = \pi \nu_0JS/2 = 0.030$, $\beta = \pi \nu_0 U = 0.011$). \textbf{e)} Topography and dI/dV for the Mn$_{18}$ chain. \textbf{f)} Energy of the $A^{1}$ YSR component of the central atom/s subtracting the superconducting tip and normalizing by the \bipd superconducting gap. We consider the central atom for an odd number of Mn, and the average of the 2 central atoms for an even number of Mn. The fit with an exponential function results in an energy saturation at $E_0=0.54\Delta$.}
		\label{F4}
  \end{center}
		\end{figure*}

To rationalize these observations, we performed model simulations using the Green Functions formalism 
 of classical spins on a superconductor, fixing their relative orientation to fit the experimental observations. The exchange coupling $J$ between impurity's spin and superconductor is extracted by fitting first the YSR energy of an isolated Mn atom. We use the experimental band structure of \bipd \cite{sakano2015} as described in Fig.~\ref{F1}e and used the $p_F$ value of the S2 surface state, $p_{F} = 0.32/a_0$,  which is found to be the main responsible of YSR states in transition metals adatoms on \bipd \cite{zaldivar2020,ortuzar2022}.

The observed shift of YSR resonances towards the gap observed in the (100)  Mn$_5$ structures is reproduced only when the spins of the four closest adatoms are antiparallel aligned to the spin of the central adatom  [Fig.~\ref{F3}g]. This confirms the precluded AFM magnetic orientation between the atoms at this separation and for this alignment.  
On the contrary, the YSR splittings observed for the  \sqrttwo and \sqrtfive Mn crosses are simulated by imposing a co-linear parallel orientation between the four nearest adatoms and the central one Figs.~\ref{F3}h and \ref{F3}i.  We found that the best fit to the experimental values of the YSR splitting in each case could be obtained by slightly reducing the value of $p_{F}$ [$p_{F}^{110} = 0.21/a_0$ and $p_{F}^{120} = 0.28/a_0$]. Beyond the simplicity of our model, the smaller value of $p_{F}$ in these cases can be due to either small non-collinear components in the spin alignment \cite{ortuzar2022}, caused by the large spin-orbit coupling in this system \cite{mier2021a}, or to the fact that the Fermi surface is not strictly squared, as depicted in Fig.~\ref{F1}e.  From these results, we demonstrate that the accumulation of YSR states from several atoms in an anisotropic superconductor induces a monotonic increase in the YSR splitting energy, as well as confirming the crucial role of the band's anisotropy: aligning the atomic distances outside the high-symmetry directions is important for inducing ferromagnetic-like spin coupling, resulting the (120) direction the one with the largest YSR split in the spectra. 

Besides the energy splitting of the \sqrttwo and \sqrtfive structures, the \sqrtfive Mn cross additionally develops an increasing particle-hole asymmetry \cite{farinacci2018,rubio-verdu2021,kamlapure2021,chatzopoulos2021} with the number of neighbors, which cannot be detected in the \sqrttwo Mn$_5$ structure. To account for this effect, we included a finite potential scattering term $\beta = \pi \nu_0 U = 0.011$ in the simulations of all the atomic structures in Fig.~\ref{F3}. The results reproduce the increase in particle-hole asymmetry for the \sqrtfive and 2a structures [Fig.~\ref{F3}g and~\ref{F3}i] and the absence of it in the \sqrttwo. As described in the appendix [Fig.~\ref{asymmetry}], the particle-hole asymmetry for an adatom increases with the YSR energy and with the strength of the YSR coupling between two interacting atoms, which is larger in the \sqrtfive than in the \sqrttwo structures due to their smaller spacing.  Engineering the particle-hole asymmetry in an adatom by bringing neighbors into its proximity can be a promising strategy for tuning the diode effect in single-atom Josephson junctions \cite{trahms2023a}.

\section{Evolution of YSR Coupling In Atomic Chains Along The (120) Direction}

Since the results found for the Mn$_5$ structures point that atoms aligned along the (120) direction exhibit the largest YSR splitting, we explore next the evolution of the observed YSR split with the size of the structure. For this purpose, we have built linear atomic chains along the (120) direction and studied the evolution of their YSR spectra with the length. 
In Fig.~\ref{F4},  we present topography and dI/dV spectroscopy of three snapshots (Mn$_3$, Mn$_9$ and Mn$_{18}$) in the construction of a (120) chain of  18 atoms. 
The Mn$_3$ chain exhibits split YSR states mostly on its central atom [Fig.~\ref{F4}a,b], similar to the \sqrtfive structures from the previous section. From this point on, adding additional adatoms to the end of the chain slightly opens up the YSR split [Fig.~\ref{F4}c,d], with increasing binding energy of the \oneysr component, until it saturates at a value of about $E_{A^1}$=0.5$\Delta$. This value is reached in chains of 6 atoms long and remains constant for longer chains [Fig.~\ref{F4}f]. 
The bandwidth saturation to  0.54$\Delta$ results from the moderate exchange $J$ of the Mn atoms with the surface electrons on this substrate. Our simulations reveal that increasing $J$ brings this atomic chain into the strong coupling regime ($A^1$ binding energy larger than $\Delta$) in a similar way as when reducing their interatomic distance \cite{mier2021a} or when enabling direct atomic exchange interaction mechanisms \cite{schneider2022}.
 
The dI/dV spectral line profile of the 9-atom and 18-atom chains, shown in Figs.~\ref{F4}c and \ref{F4}e, also reveals a homogeneous distribution in the energy and amplitude of the YSR peaks along the chain, but only a continuous shift to lower \oneysr binding energies for the last 3 Mn atoms. Such flat-band-like behavior is also a characteristic of a weak atom-surface interacting system. Increasing the value of $J$ results in a more significant energy dispersion of YSR states and the emergence of quantum-well-like states in the simulations \cite{green-ysr}. The gradual shift of \oneysr toward the gap edge at the chain ends indicates that each atom interacts only with the three nearest neighbors at each side.  This explains the stabilization of the \oneysr binding energy of the central atoms for chains larger than six atoms long. 
 
\begin{figure}[t]
\begin{centering}
			\includegraphics[width=1\columnwidth]{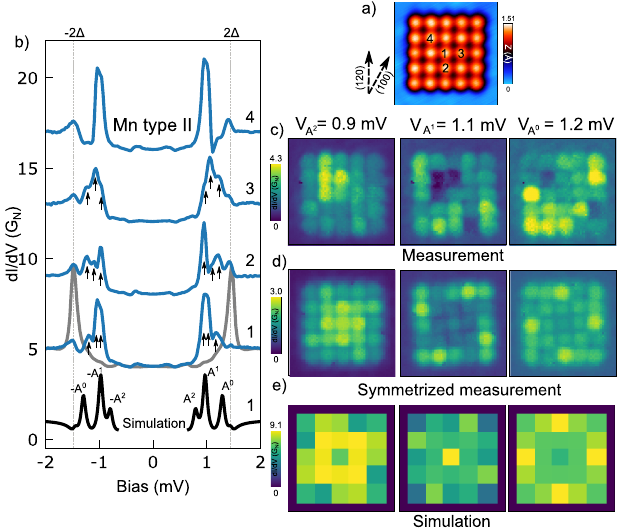}    
			\caption{\textbf{Mn$_{25}$ array along \sqrtfivedir.} \textbf{a)} Topography image of the 25 Mn atoms squared array (4.2 x 4.2 nm, I = 20 pA, I = 40 mV). \textbf{b)} Spectra on indexed atoms in a) (1,2,3,4), vertical offset for clarity. The YSR resonances ($A^{0}$, $A^{1}$, and $A^{2}$) are highlighted in the spectra identified by the black markers and labeled in the simulation of the central atom at the bottom (black curve). The spectrum labeled 4 is of a type II Mn atom, showing the characteristic energy dip [Fig.~\ref{statistics}a in Appendix] (I = 400 pA, V = 4mV). \textbf{b)} Energy cuts of a 64x64 dI/dV grid measured on the squared atom array at the energy of the three resonances highlighted in the spectra. \textbf{d)} Symmetrized experimental map obtained by rotating 90° and averaging the maps in c). \textbf{e)} Green function simulations at the three resonance energies including the Mn type II in position 4 (simulation parameters $p_{F1} = 0.28/a_0$, $m^* = 5.85/m_e$, $\alpha = \pi \nu_0JS/2 = 0.030$, $\alpha_{\text{Mn type2}} = \pi \nu_0J^*S/2 = 0.06$, $\beta = \pi \nu_0 U=0$).}
		\label{F5}
  \end{centering}
		\end{figure}

\section{Mapping of the collective YSR modes in an array of 25 Mn atom}
The results of the Mn chains built along the (120) direction of \bipd reveal that the YSR binding energy at the center of the structure increases initially and remains constant for chains longer than 6 atoms. A possible route to enhance the effect of YSR overlap is extending the array in the 2D plane. To explore the impact of dimensionality, we have built a 5x5 Mn squared array of Mn adatoms along the (120) substrate directions, with \sqrtfive lattice spacing [Fig.~\ref{F5}a]. Spectra inside the Mn$_{25}$ island show a strong dependence on the specific atomic site. For example, in Fig.~\ref{F5}b, we compare the differential conductance spectra measured on four atoms of the structure labeled as 1,2,3,4 in Fig.~\ref{F5}a. Overall, we identify three main in-gap spectroscopic features in the structure, indicated by arrows in their spectra in Fig.~\ref{F5}b. The appearance of multiple YSR peaks and the larger binding energy for the first of the YSR resonances (appearing at $\epsilon \approx \pm 0.9$~mV in the spectra, i.e. 0.2$\Delta$) hints at the more considerable YSR overlap in these structures.

To identify the origin of the modes in panel Fig.\ref{F5}b, we map their spatial distribution \cite{soldini2023a} with a spectral grid over the Mn$_{25}$ structure. 
Figure~\ref{F5}c shows constant bias dI/dV maps extracted from the grid at the energies of the three in-gap features $V_{A^{2}} \sim 0.9$ mV, $V_{A^{1}} \sim 1.1$ mV, $V_{A^{0}} \sim 1.2$ mV. Contrary to our expectations, the maps are not symmetric due to an inhomogeneity in atom 4. 
Over this atom, the dI/dV spectrum shows a YSR peak at slightly larger binding energy, as occasionally observed on some of the Mn atoms investigated (type II Mn atoms in Fig.~\ref{statistics} in the Appendix), probably due to a subsurface vacancy. 
The different YSR binding energy at this adatom is probably the cause of the lack of a fourfold symmetry in the maps of the 5x5 square array. 
To recover the C$_4$ symmetry,  we tentatively averaged the dI/dV maps with their three 90° rotations (see Appendix \ref{app:asymm} for further details). By this procedure, we intend to enhance the symmetric LDOS features over the defect-induced ones and wash out the effect caused by the defect. The symmetrized maps, shown in Fig.~\ref{F5}d, show now that the state $A^{2}$ is localized mainly over the center of the array, the $A^{1}$ mostly over the atoms close to the corners, and the state $A^{0}$ at the edges of the square.

To correlate these features with the ideal LDOS of an unperturbed atomic array, we simulated the 25-atom structure using, as before, parameters obtained from the YSR spectra of the isolated Mn atoms. The plot at the bottom of Fig.~\ref{F5}b (black line) shows the simulated LDOS spectra over the central atom (adatom 1), which reproduces the three in-gap YSR states found in the experiment. Tuning the model parameters, we find that the number of spectral peaks increases with the exchange coupling $J$ value and the closer distance between adatoms, analogous to quantum well states \cite{schneider2022,mier2021}. 

The spatial distribution of these three resonances, shown in Fig.~\ref{F5}e, shows a four-fold symmetric spatial distribution that agrees with the C$_4$ symmetry of the Mn$_{25}$ island. The simulated maps can also be correlated with the general features of the symmetrized experimental maps, such as localization of LDOS of the peaks $A^{2}$, $A^{1}$ and $A^{0}$  over the center and edges, respectively.  
The simulated maps also reproduce the absence of mirror-symmetry planes and the weak chirality of the symmetrized dI/dV maps. We note that the C$_4$ Mn atomic structure is rotated by $\approx$ 26 degrees concerning the \bipd lattice and, therefore, does not have mirror planes when the surface is also considered but, instead, has a chiral structure. 
The chirality of the adsorbate-substrate structure can be reflected in their normal state LDOS \cite{Mugarza10}. Here, we show it is also present in subgap YSR spectral features. Remarkably, the chiral structure of the adatom array can be resolved by symmetrizing dI/dV maps, revealing that it persists over the distortion caused by a single adatom defect. 

\section{CONCLUSIONS}

In conclusion, we have shown that Mn atomic arrays on the surface of the superconductor \bipd exhibit interacting YSR states in the diluted limit. In this article, we have studied the evolution of their YSR spectra and their spatial distribution with the orientation and size of the atomic structures. Owing to this material's highly anisotropic band structure, we found that the type and size of the indirect magnetic coupling between adatoms strongly depends on their relative orientation, besides their separation. 
We performed Greens Function simulations of the subgap LDOS to interpret our observations, including the characteristic square Fermi contours of \bipd  as detailed \cite{ortuzar2022}. Specifically, we used the code ``green-ysr'', which computes arbitrary lattices of magnetic adatoms coupled to superconductors with anisotropic Fermi surfaces and is available online \cite{green-ysr}. The simulations reproduce the anisotropic YSR coupling between magnetic adatoms in the dilute limit. 
We found that Mn structures along the (120) surface direction exhibit the most significant coupling with ferromagnetic components, detected as a split of YSR peaks in subgap spectra. In atomic chains engineered along this direction, the YSR split increases with the size until chains of more than six atoms remain constant, indicating that the interaction of an adatom is restricted to its 3-4 neighboring atoms. We also extended the arrays to two dimensions by building a squared array of 25 atoms and observed their collective YSR modes with a peculiar spatial distribution. Supported by the simulations, we identified that the maps reflect a chiral LDOS, probably originated by the rotation of the squared Mn lattice rotated over the \bipd lattice. Furthermore, we found that the anisotropic character of the Fermi surface offers a novel route to tune YSR energy and particle-hole asymmetry of a diluted adatoms array.

\begin{acknowledgments}
We acknowledge financial support from the Spanish \\ MCIN/AEI/10.13039/501100011033 and the European Regional Development Fund (ERDF) through grants
PID2020-114071RB-I00, PID2020-114252GB-I00, PID2022-140845OB-C61, 
CEX2020-001038-M, CEX2023-001316-M, 
TED2021-130292B-C42,
and TED2021-130546B-I00, 
the Basque Government through grant IT-1591-22, and the Comunidad de Madrid through grant  NANOMAGCOST-CM (Program No.S2018/NMT-4321). We also acknowledge collaborations through EU program Cost CA21144 (Superqumap), and the ERC-AdG CONSPIRA (101097693).  
J.O. acknowledges the scholarship PRE-2021-1-0350 from the Basque Government. 
\end{acknowledgments}

\appendix

\counterwithin{figure}{section}

\section{The Theoretical Model}\label{app:theo}

In this appendix, we describe the analytical model applied to simulate the data presented in the main text. For an extended explanation, refer to  \citet{ortuzar2022}. 

This work presents the effects of the interaction between a diverse variety of magnetic impurity structures. These structures are assumed to be in the diluted impurity limit, meaning that the itinerant electrons mainly mediate the interaction between impurities. This interaction can be described at the second-order approximation as a RKKY interaction. Other works have used this fact to postulate effective tight-binding Hamiltonian to describe systems with many impurities \cite{liebhaber2022}. Our model diverges from this approach, as we have calculated the exact Green Functions for the complex superconductor-impurity system. The advantage of our approach is that we can calculate the decay of the YSR wave function as well as its anisotropic behavior due to the non-circular shape of the Fermi contour of the \bipdp.

We calculate the LDOS of interacting magnetic impurities from the real space Green Function. Assuming that the chemical potential with respect to the band bottom is greater than the superconducting gap ($\mu\gg\Delta$) one can perform analytically the Fourier transform of the momentum space Green Functions of a superconductor with arbitrary geometrical shaped Fermi contour. Once this is known, we calculate the perturbed Green Function utilizing the Dyson formula \cite{abrikosov2012}:
\begin{equation}
    G(r,r')=G_0(r-r')+\sum_n G_0(r-r_n)V_nG(r_n,r')
\end{equation}
$G$ and $G_0$ are the perturbed and bare Green Functions, respectively, $V_n$ is the potential of the nth magnetic impurity and $r_n$ its position. This can be exactly calculated \cite{balatsky2006}, and $G(r-r')$ can be described as a function of $G_0(r-r')$, which we can analytically calculate. 

The final formula will depend on a set of parameters that we utilize in the main text, namely: the Fermi momentum ($p_F$) and superconducting gap of the surface ($\Delta$), the exchange coupling of the impurities with the surface ($\alpha_n\propto V_n$) and the position of the impurities ($r_n$). With this set of parameters ($\{p_F,\Delta,\alpha_n,r_n\}$), we can calculate the LDOS at any point and energy. Most of the parameters are easily obtainable from the isolated impurity spectrum. The relative positions of the impurities are acquired from the topographic images, and their coupling from the YSR energies, utilizing the known energy dispersion $\epsilon_{YSR}=\pm\Delta(1-\alpha^2)(1+\alpha^2)^{-1}$. The superconducting gap can be obtained from spectroscopic measurements on the bare superconductor. Finally, for $p_F$, although we mainly use the Fermi momenta from the literature \cite{herrera2015}, the approximated perfect square-like Fermi contour we use for the calculations can overestimate the size of the contour compared with the real one. For this reason, we adjusted the Fermi momentum depending on the direction of the sample, as explained in the main text [Fig.~\ref{F2}e].

\section{Statistical analysis of YSR states energy}\label{app:stat}

\begin{figure}[b]
        \begin{center}
			\includegraphics[width=1\columnwidth]{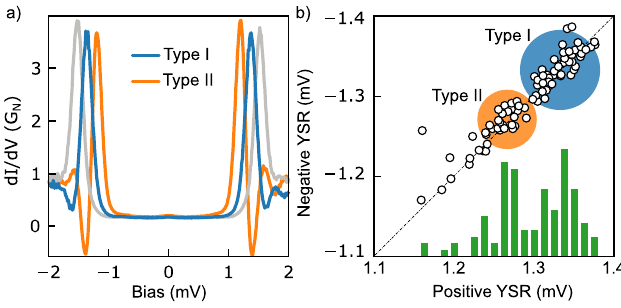}
        \end{center}
			\caption{\textbf{a)} YSR spectra acquired for two Mn atoms on two types of adsorption sites of \bipdp. The type II adsorption site presents lower YSR energy and a more pronounced negative differential conductance (V = 3 mV, I = 300 pA). \textbf{b)} Statistical distribution of YSR energies extracted from 103 spectra using the same superconducting tip. The histogram with 0.125 $\mu$V bin shows accumulation in categories I and II, highlighted by the orange and blue circles.}
		\label{statistics}
	\end{figure}

The \bipd superconducting substrate presents Bi atoms vacancies in the 1st layer and Bi adatoms on the surface. The deposited Mn atoms considered in this study are always sufficiently far from 1st layer vacancies, but we cannot exclude an under-layer vacancy that can affect the YSR energy. Indeed, the YSR energy presents some small modulations that we quantify statistically in Fig.~\ref{statistics}a. We show that negative and positive YSR components are energy-symmetric but with a distribution peaked at two energy ranges. Differential conductance spectra for the two cases [Fig.~\ref{statistics}b] show a difference in the YSR resonance energy position and a pronounced negative differential conductance (NDC) due to the convolution with the superconducting tip. The two different types of Mn are due to surface effects, like an under-surface vacancy. In fact, the same atom can be manipulated between different sites and show characteristics of type I or type II. Before constructing an array, we check the type of each lattice site by manipulating one Mn atom and measuring its spectrum.

\section{Particle-hole asymmetry in Mn$_5$ clusters}\label{app:asymm}

\begin{figure}[h]
        \begin{center}
			\includegraphics[width=0.5\textwidth]{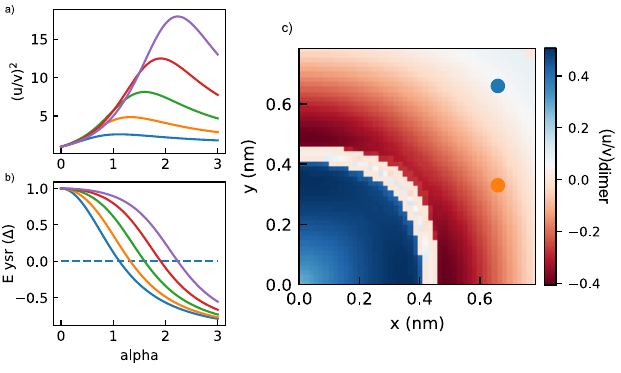}
        \end{center}
			\caption{\textbf{a)} Particle hole asymmetry for different values of potential scattering ($\beta$ = 0.5, 0.875, 1.25, 1.625, 2) of the YSR resonances depending on $\alpha$ (adimensional exchange). \textbf{b)} YSR energy dependence on $\alpha$ for the same value of the potential scattering of b). \textbf{c)} Simulation with the Green function model for a dimer with the first atom located in the origin (simulation parameters  $p_{F}^{110} = 0.21/a_0$, $m^* = 5.85/m_e$, $\alpha = \pi \nu_0JS/2 = 0.030$, $\beta = \pi \nu_0 U = 0.011$). We plot the calculated normalized particle-hole asymmetry $(u/v)_{dimer}$ and mark the \sqrtfive (orange) and \sqrttwo (blue) positions for the second atom. Note that when the atoms are sufficiently close the particle-hole asymmetry changes sign, indicating the quantum phase transition.}
		\label{asymmetry}
	\end{figure}

Apart from YSR hybridization, a pronounced asymmetry is frequently observed in extended YSR lattices \cite{schneider2022,soldini2023a}. In the 5 Mn atoms cluster, we observe an increase in the particle-hole asymmetry as we add more atoms. We shown in section III of the main text that our model captures the particle-hole asymmetry by introducing a potential scattering term in the simulation. In general, the particle-hole asymmetry ($u/v$)$^2$ for a classical impurity follows the expression \cite{huang2021}:
\begin{equation}
    \bigg(\frac{u^2}{v^2}\bigg)_{monomer} = \frac{1+(\alpha+\beta)^2}{1+(\alpha-\beta)^2}.
\end{equation}
In Fig.~\ref{asymmetry}a we plot ($u/v$)$^2_{monomer}$ in the function of alpha for different values of the potential scattering parameter. Before the quantum phase transition (QPT) (visible in Fig.~\ref{asymmetry}b) there is an increase in particle-hole asymmetry. For a dimer, the asymmetry depends on both $\alpha$ and the inter-atomic distance, which modulates the YSR energy. We compute this with our model for a dimer in Fig.~\ref{asymmetry}c, graphing the following ratio:
\begin{equation}
    \bigg(\frac{u}{v}\bigg)_{dimer} = \dfrac{\rho^{A^1} - \rho^{-A^1}} {\rho^{A^1}+\rho^{-A^1}},
\end{equation}
where $\rho$ is the calculated conductance of the inner YSR component $A^1$. We see that $(u/v)_{dimer}$ strongly depends on the inter-atomic distance and the position of the YSR components. The asymmetry in the \sqrttwo atomic position (blue dot) is minimum and finite in the \sqrtfive position (orange dot), matching the larger particle-hole asymmetry increase in the \sqrtfive Mn$_5$ cluster [Fig.~\ref{F3}]. Like the YSR splitting, also the particle-hole asymmetry changes with the inter-atomic distance, thus allowing for its engineering.

\section{Additional simulations of the 25 Mn atom array}\label{app:grid}

\begin{figure}[ht]
        \begin{center}
			\includegraphics[width=0.95\columnwidth]{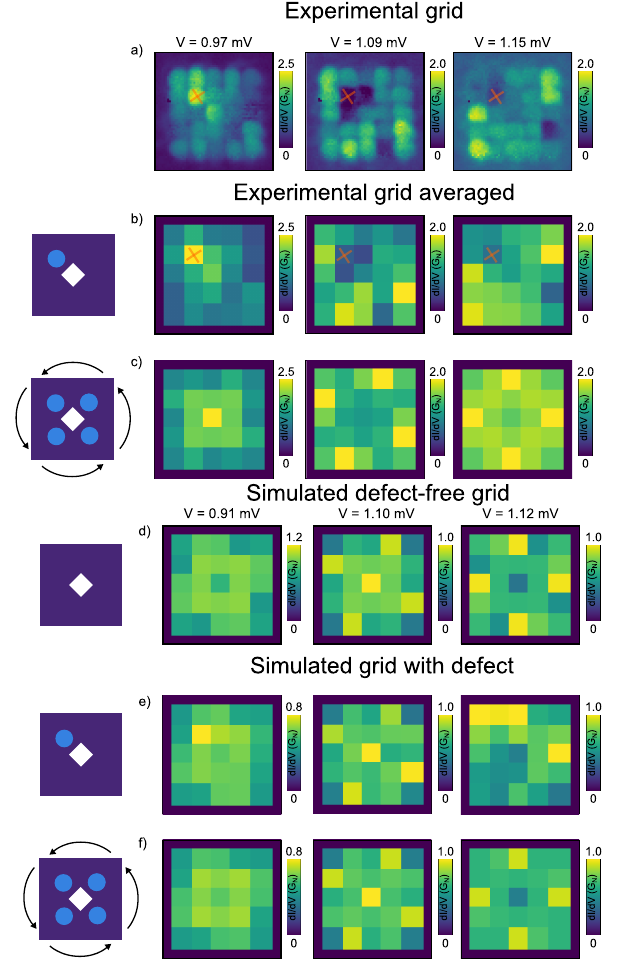}
        \end{center}
			\caption{\textbf{a)} Constant energy cuts at the YSR modes energies of the experimental dI/dV raw grid of the 25 Mn atom square structure presented in Fig 5 of the main text. The orange cross indicates the type II Mn atom. \textbf{b)} Cuts at the YSR modes energies of the averaged experimental dI/dV grid of a), averaging over the atomic positions to better compare with the simulation. \textbf{c)} Cuts at the YSR modes energies of the averaged and symmetrized experimental dI/dV grid using the symmetrization procedure described in the text. \textbf{d)} Cuts at the YSR energies of the defect-free simulation of the 25 Mn atom square structure, same parameters as reported in the main text. \textbf{e)} Simulated grid with a defect, visible as an enhanced density of states in the first cut. \textbf{f)} Cuts of the simulated grid with a defect, showing similarity with the defect-free grid. (simulation parameters $p_{F1} = 0.28/a_0$, $m^* = 5.85/m_e$, $\alpha = \pi \nu_0JS/2 = 0.030$, $\alpha_{\text{Mn type2}} = \pi \nu_0J^*S/2 = 0.06$, $\beta = \pi \nu_0 U=0$)}
		\label{additional_grids}
	\end{figure}

In Fig.~\ref{additional_grids} we report a series of simulation of the 25 Mn atom array of Fig.~\ref{F5}. In Fig.~\ref{additional_grids}a we reproduce from the main text the experimental dI/dV cuts of the 64x64 spectral grid, that we average around the atomic positions in Fig.~\ref{additional_grids}b. This maps are than symmetrized by averaging four 90-degree rotations Fig.~\ref{additional_grids}c. This experimental data can be simulated in two different ways: with a perfect array of type I Mn atoms [Fig.~\ref{additional_grids}d] or with an array containing one type II Mn atom [Fig.~\ref{additional_grids}e] and applying the symmetrization procedure [Fig.~\ref{additional_grids}f]. The resulting simulated maps of Fig.~\ref{additional_grids}d and \ref{additional_grids}f show the same YSR modes spatial distribution. This confirms that the symmetrization procedure enhances the symmetric LDOS components over the defect-induced ones without interfering with the YSR modes, that appear in the defect-free simulation.

\newpage

\end{document}